\documentclass[useAMS,usegraphicx]{mn2e}

\newcommand{\etal}{et al.}
\newcommand{\rxjw}{RX~J1856.5$-$3754}
\newcommand{\zg}{z_g}
\newcommand{\ycol}{y_{\rm H}}
\newcommand{\lambatm}{\lambda_{\rm thin}}
\newcommand{\tatm}{T_{\rm thin}}
\newcommand{\taulam}{\tau_\lambda}
\newcommand{\nh}{\ensuremath{N_{\rm H}}}
\newcommand{\teff}{\ensuremath{T_{\rm eff}}}
\newcommand{\rem}{\ensuremath{R^{\rm em}}}
\newcommand{\ThetaB}{\ensuremath{\Theta_B}}
\newcommand{\tinfty}{\ensuremath{T^{\infty}}}
\newcommand{\rinfty}{\ensuremath{R^{\infty}}}
\newcommand{\xmm}{\textit{XMM-Newton}}
\newcommand{\chandra}{\textit{Chandra}}
\newcommand{\hst}{\textit{HST}}
\newcommand{\euve}{\textit{EUVE}}
\newcommand{\expnt}[2]{\ensuremath{#1 \times 10^{#2}}}   
\newcommand{\mc}{\multicolumn}

\title[Hydrogen Atmospheres and \rxjw]{Magnetic Hydrogen Atmosphere Models
and the Neutron Star \rxjw}

\author[Ho, Kaplan, Chang, van Adelsberg, \& Potekhin]{Wynn C. G.
Ho$^{1,2}$\thanks{Hubble Fellow},
David L. Kaplan$^{1}$\thanks{Pappalardo Fellow},
Philip Chang$^{3,4}$,
\newauthor
Matthew van Adelsberg$^{5}$,
and Alexander Y. Potekhin$^{5,6}$ \\
$^{1}$Kavli Institute for Astrophysics and Space Research,
Massachusetts Institute of Technology, Cambridge, MA, 02139, USA \\
$^{2}$Kavli Institute for Particle Astrophysics and Cosmology,
Stanford University, P.O. Box 20450, Mail Stop 29, Stanford, CA, 94309, USA \\
$^{3}$Department of Astronomy, 601 Campbell Hall, University of California,
Berkeley, CA 94720, USA \\
$^{4}$Department of Physics, Broida Hall, University of California,
Santa Barbara, CA, 93106, USA \\
$^{5}$Center for Radiophysics and Space Research, Department of Astronomy,
Cornell University, Ithaca, NY, 14853, USA \\
$^{6}$Ioffe Physico-Technical Institute, Politekhnicheskaya 26,
194021 St Petersburg, Russia \\
}

\begin{document}

\date{Accepted . Received ; in original form}

\pagerange{\pageref{firstpage}--\pageref{lastpage}} \pubyear{2006}

\maketitle

\label{firstpage}

\begin{abstract}
\rxjw\ is one of the brightest nearby isolated neutron stars,
and considerable observational resources have been devoted to it.
However, current models are unable to satisfactorily explain the data.
We show that our latest models of a thin, magnetic, partially ionized hydrogen
atmosphere on top of a condensed surface can fit the entire spectrum,
from X-rays to optical, of \rxjw, within the uncertainties.
In our simplest model, the best-fit parameters are
an interstellar column density $\nh\approx 1\times 10^{20}\mbox{cm$^{-2}$}$
and an emitting area with $\rinfty\approx 17$~km (assuming
a distance to \rxjw\ of 140~pc), temperature $\tinfty\approx 4.3\times
10^5$~K, gravitational redshift $\zg\sim 0.22$, atmospheric hydrogen
column $\ycol\approx 1\mbox{ g cm$^{-2}$}$, and magnetic field
$B\approx (3-4)\times 10^{12}$~G;
the values for the temperature and magnetic field indicate an
effective average over the surface.  We also calculate a more realistic
model, which accounts for magnetic field and temperature variations
over the neutron star surface as well as general relativistic effects,
to determine pulsations;  we find there exist viewing geometries that
produce pulsations near the currently observed limits.
The origin of the thin atmospheres required to fit the data is an
important question, and we briefly discuss mechanisms for producing
these atmospheres.  Our model thus represents the most self-consistent
picture to date for explaining all the observations of \rxjw.
\end{abstract}


\section{Introduction} \label{sec:intro}

Seven candidate isolated, cooling neutron stars (INSs) have
been identified by the ROSAT All-Sky Survey, of which the two brightest
are \rxjw\ and RX~J0720.4$-$3125
(see Treves \etal~2000; Pavlov, Zavlin, \& Sanwal~2002;
Kaspi, Roberts, \& Harding~2006
for a review).  These objects share the following properties: (1) high
X-ray to optical flux ratios of $\log(f_{\rm X}/f_{\rm optical})\sim 4-5.5$,
(2) soft X-ray spectra that are well described by blackbodies with
$kT\sim 50-100$~eV, (3) relatively steady X-ray flux over long timescales,
and (4) lack of radio pulsations.

For the particular INS \rxjw,
single temperature blackbody fits to the X-ray spectra
underpredict the optical flux by a factor of $\sim 6-7$
(see Fig.~\ref{fig:sp}).
X-ray and optical/UV data can best be fit by two-temperature
blackbody models with
$k\tinfty_{\rm X}=63$~eV, emission size
$\rinfty_{\rm X}=5.1\,(d/\mbox{140 pc})$~km,\footnote{The most
recent determination of the distance to \rxjw\ is $\approx 160$~pc
(Kaplan \etal, in preparation).  However, the uncertainties in this
determination are still being examined.  Therefore, we continue to use
the previous estimate of $140 (\pm 40)$~pc from Kaplan, van Kerkwijk,
\& Anderson~(2002) since the uncertainty in this previous value
encompasses both the alternative estimate of 120~pc from
Walter \& Lattimer~(2002) and the new value. \label{foot:distance}}
$k\tinfty_{\rm opt}=26$~eV, and
$\rinfty_{\rm opt}=21.2\,(d/\mbox{140 pc})$~km
(Burwitz \etal~2001, 2003; van Kerkwijk \& Kulkarni~2001a;
Braje \& Romani~2002; Drake \etal~2002; Pons \etal~2002;
see also Pavlov \etal~2002; Tr\"{u}mper \etal~2004),
where $\tinfty=\teff/(1+\zg)$, $\rinfty=\rem(1+\zg)$,
and $\rem$ is the physical size of the emission region.
The gravitational redshift $\zg$ is given by $(1+\zg)=(1-2GM/Rc^2)^{-1/2}$,
where $M$ and $R$ are the mass and radius of the NS, respectively.
However, the lack of X-ray pulsations (down to the 1.3\% level)
puts severe constraints on such two-temperature models
(Drake \etal~2002; Ransom, Gaensler, \& Slane~2002; Burwitz \etal~2003).
It is possible that the magnetic axis is aligned with the
spin axis or the hot magnetic pole does not cross our line of sight
(Braje \& Romani~2002).  Alternatively, \rxjw\ may possess a superstrong
magnetic field ($B\ga 10^{14}$~G) and has spun down to a period $> 10^4$~s
(Mori \& Ruderman~2003),
though Toropina, Romanova, \& Lovelace~(2006) argue that this last case
cannot explain the H$\alpha$ nebula found around \rxjw\
(van Kerkwijk \& Kulkarni~2001b).
On the other hand, a single uniform temperature is possible
if the field is not dipolar but small-scale (perhaps due to turbulence
at the birth of the NS; see, e.g., Bonanno, Urpin, \& Belvedere~2005,
and references therein).

Even though blackbody spectra fit the data, one expects NSs to possess
atmospheres of either heavy elements (due to debris from the progenitor)
or light elements (due to gravitational settling or accretion);
we note that a magnetized hydrogen atmosphere may provide a consistent
explanation for the broad spectral feature seen in the atmosphere of
RX~J0720.4$-$3125 (Haberl \etal~2004; Kaplan \& van~Kerkwijk 2005).
The lack of any significant spectral features in
the X-ray spectrum argues against a heavy element atmosphere (Burwitz
\etal~2001, 2003), whereas single
temperature hydrogen atmosphere fits overpredict the optical flux by a
factor of $\sim 100$ (Pavlov \etal~1996; Pons \etal~2002; Burwitz
\etal~2003).  However, these hydrogen atmosphere results are derived
using non-magnetic atmosphere models.  Only a few magnetic (fully
ionized) hydrogen or iron atmospheres have been considered (e.g.,
Burwitz \etal~2001, 2003), and even these models are not adequate.
Since $kT\sim\mbox{tens of}$~eV for
\rxjw\ and the ionization energy of hydrogen at $B=10^{12}$~G is
160~eV, the presence of neutral atoms must be accounted for in the
magnetic hydrogen atmosphere models; the opacities are sufficiently
different from the fully ionized opacities that they can change the
atmosphere structure and continuum flux (Ho \etal~2003; Potekhin \etal~2004),
which can affect fitting of the observed spectra.

Another complication in fitting the observational data of \rxjw\
(and RX~J0720.4$-$3125) with hydrogen atmosphere models is that the model
spectra are harder at high X-ray energies.
On the other hand, observations of RX~J0720.4$-$3125
suggest it possesses a dipole
magnetic field $B\approx 2\times 10^{13}$~G (Kaplan \& van Kerkwijk 2005).
It is probable then that RX~J0720.4$-$3125 (and possibly \rxjw)
is strongly magnetized with $B\sim 10^{13}-10^{14}$~G,
and its high energy emission is softened by the effect of vacuum
polarization, which can show steeper high energy tails
(Ho \& Lai~2003).
Rather than resorting to a superstrong magnetic field,
an alternate possibility is that there exists a ``suppression'' of the high
energy emission.  One such mechanism is examined in
Motch, Zavlin, \& Haberl~(2003; see also Zane, Turolla, \& Drake~2004),
specifically, a geometrically thin hydrogen atmosphere at the surface
that is optically thick to low energy photons and optically thin to
high energy photons.
The high energy photons that emerge then bear the signature of the
lower temperature (compared to atmospheres that are
optically thick at all energies) at the inner boundary layer
(usually taken to be a blackbody) of the atmosphere model;
this leads to a softer high energy tail.
Motch \etal~(2003) find a good fit to RX~J0720.4$-$3125 in this case
by using a non-magnetic atmosphere model with $k\teff=57$~eV,
a hydrogen column density $\ycol=0.16\mbox{ g cm$^{-2}$}$,
and distance of 204~pc, and assuming a $M=1.4 M_\odot$, $R=10$~km NS.

We examine this last possibility by fitting the entire spectrum
of \rxjw\ with the latest partially ionized hydrogen atmosphere models
(constructed using the opacity and equation of state tables from
Potekhin \& Chabrier~2003) and condensed matter in strong magnetic fields
(see van Adelsberg \etal~2005).
The goal of the paper is to provide a self-consistent picture of \rxjw\
that resolves the major observational and theoretical inconsistencies:
(1) blackbodies fit the spectrum much better than realistic atmosphere models,
(2) strong upper limits on X-ray pulsations suggest \rxjw\ may have a largely
uniform temperature (and hence magnetic field) over the entire NS surface,
(3) the inferred emission size from blackbody fits are either much
smaller or much larger than the canonical NS radius of $10-12$~km.
Because of observational uncertainties  (see Section~\ref{sec:obs}) and
the computationally tedious task of constructing a complete
grid of models, we do not attempt to prove the uniqueness of our results;
rather we try only to reproduce the overall spectral energy distribution
and argue for the plausibility of our model.

An outline of the paper is as follows.
In Section~\ref{sec:atm}, we describe the atmosphere model
(a thin, magnetized atmosphere of partially ionized hydrogen with a
condensed surface) and possible ways of producing this atmosphere.
A brief description of the observations is given in Section~\ref{sec:obs}.
The model fits to the observational data are shown in Section~\ref{sec:atmfit}.
Possible mechanisms for the creation of thin hydrogen atmospheres and
pulsations implied by our model are given in Section~\ref{sec:discussion}.
We summarize and discuss our results in Section~\ref{sec:conclusion}.

\section{Models of Neutron Star Atmospheres}
\label{sec:atm}
Thermal radiation from a NS is mediated by the atmosphere
(with scaleheight $\sim 1$~cm).
In the presence of magnetic fields typical of INSs
($B\ga 10^{12}$~G), radiation propagates in two polarization modes
(see, e.g., M\'{e}sz\'{a}ros~1992).
Therefore, to determine the emission properties of a magnetic atmosphere,
the radiative transfer equations for the two coupled photon polarization
modes are solved.
The self-consistency of the atmosphere model is determined by requiring
the fractional temperature corrections $\Delta T(\tau)/T(\tau)\ll 1\%$
at each Thomson depth $\tau$, deviations from radiative
equilibrium $\ll 1\%$, and deviations from constant flux
$< 1\%$ (see Ho \& Lai~2001; Ho \etal~2003; Potekhin \etal~2004, and
references therein for details on the construction of the atmosphere models).
We note that the atmosphere models formally have a dependence, through
hydrostatic balance, on the surface gravity $g$~$[=(1+\zg)GM/R^2]$
and thus the NS radius $R$; however, the emergent spectra
do not vary much using different values of $g$ around
$2\times 10^{14}$~cm~s$^{-2}$ (Pavlov \etal~1995).
Nevertheless, we construct models using a surface gravity that is
consistent with the inferred radius obtained from the spectral fits
in Section~\ref{sec:atmfit}
($g=1.1\times 10^{14}$~cm~s$^{-2}$ with $M=1.4M_\odot$, $R=14$~km,
and $\zg=0.2$).
Also, though our atmosphere models can have a magnetic field at an
arbitrary angle $\ThetaB$ relative to the surface normal, the models
considered in Sections~\ref{sec:atm} and \ref{sec:atmfit} have the
magnetic field aligned perpendicular to the stellar surface
(see Section~\ref{sec:pulse} for justification and other cases).
We describe other elements of our atmosphere models below.

\subsection{Partially Ionized Atmospheres}
\label{sec:atm_pi}

As discussed in Section~\ref{sec:intro}, previous works that attempted
to fit the spectra of \rxjw\ and other INSs with magnetic hydrogen
atmosphere models assume the hydrogen is fully ionized.
The temperature obtained using these models (or simple
blackbodies) are in the range $k\tinfty\approx 40-110$~eV.
Contrast this with the atomic hydrogen binding energies of 160~eV and
310~eV at $B=10^{12}$~G and $10^{13}$~G, respectively.
Therefore the atmospheric plasma must be partially ionized.

\begin{figure}
\includegraphics[width=84mm]{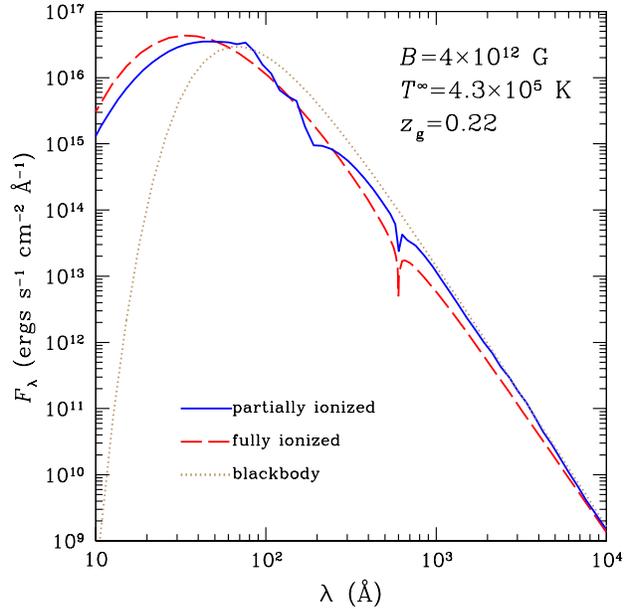}
\caption{
Spectra of hydrogen atmospheres with $B=4\times 10^{12}$~G,
and $\tinfty=4.3\times 10^5$~K.
The solid line is for a partially ionized atmosphere,
the dashed line is for a fully ionized atmosphere,
and the dotted line is for a blackbody.
All spectra are redshifted by $\zg=0.22$.
}
\label{fig:sp_pi}
\end{figure}

Figure~\ref{fig:sp_pi} illustrates the spectral differences between
a fully ionized and a partially ionized hydrogen atmosphere.
The atomic fraction is $<10\%$ throughout the atmosphere,
where the atomic fraction is the number of H atoms with non-destroyed
energy levels divided by the total number of protons
(see Potekhin \& Chabrier~2003; Ho \etal~2003).
Besides the proton cyclotron line at
$\lambda_{Bp}=1966(B/10^{12}\mbox{ G})^{-1}(1+\zg)$~\AA,
the other features are due to bound-bound and bound-free transitions.
In particular, these are the $s=0$ to $s=1$ transition at (redshifted) wavelength $\lambda=170$~\AA,
the $s=0$ to $s=2$ transition at $\lambda=110$~\AA,
and the bound-free transition at $\lambda=61$~\AA.
The quantum number $s$ measures the $B$-projection of the relative
proton-to-electron angular momentum (see, e.g., Lai~2001). For a moving
atom, this projection is not an integral of motion, but nonetheless the
quantum number $s$ (or $m=-s$) remains unambiguous and convenient for
numbering discrete states of the atom (see Potekhin 1994).
Because of magnetic broadening, the features resemble dips rather than
ordinary spectral lines (see Ho \etal~2003).

\subsection{Thin Atmospheres}
\label{sec:atm_thin}

Conventional NS atmosphere models assume the atmosphere is geometrically
thick enough such that it is optically thick at all photon energies
($\taulam\gg 1$ for all wavelengths $\lambda$, where the optical depth is
$\taulam=\int\chi_\lambda\,ds$, the extinction coefficient is
$\chi_\lambda$, and the distance traveled by the photon ray is $ds$);
thus the observed photons are all created within the atmosphere layer.
The input spectrum (usually taken to be a blackbody) at the bottom of the
atmosphere is not particularly important in determining the spectrum seen
above the atmosphere since photons produced at this innermost layer undergo
many absorptions/emissions.  The observed spectrum is
determined by the temperature profile and opacities of the atmosphere.
For example, atmosphere spectra are harder than a blackbody (at the same
temperature) at high energies as a result of the non-grey opacities
(see Fig.~\ref{fig:sp_pi}); the opacities decline with energy so that
high energy photons emerge from deeper, hotter layers in the atmosphere
than low energy photons.

On the other hand, consider an atmosphere that is geometrically thinner
than described above, such that the atmosphere is optically thin at high
energies but is still optically thick at low energies
($\taulam<1$ for $\lambda<\lambatm$ and $\taulam>1$ for $\lambda>\lambatm$).
Thus photons with wavelength $\lambda<\lambatm$ pass through the
atmosphere without much attenuation
(and their contribution to thermal balance is small since most of the
energy is emitted at $\lambda>\lambatm$ in the case of \rxjw).
If the innermost atmosphere layer (at temperature $\tatm$) emits as
a blackbody, then the observed spectrum at $\lambda<\lambatm$ will
just be a blackbody spectrum at temperature $T=\tatm$.
Motch \etal~(2003) showed that a ``thin'' atmosphere can yield a
softer high energy spectrum than a ``thick'' atmosphere
and used a thin atmosphere spectrum to fit the observations of
RX~J0720.4$-$3125.

\begin{figure}
\includegraphics[width=84mm]{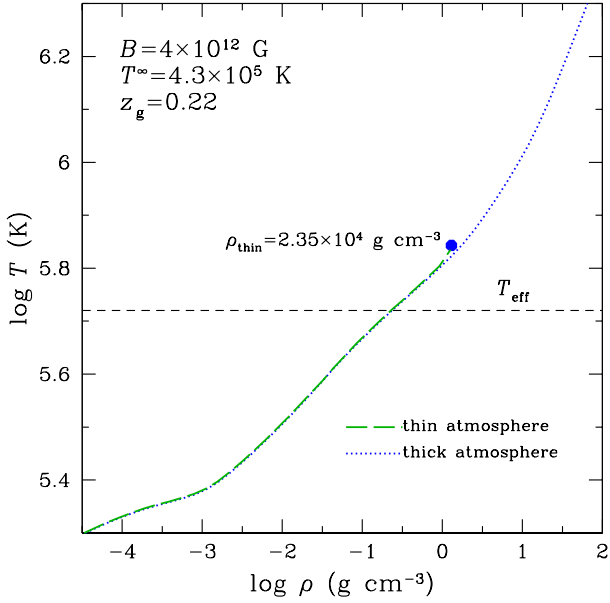}
\caption{
Model atmosphere temperature profiles with $B=4\times 10^{12}$~G and
$\tinfty=4.3\times 10^5$~K.
The dotted and long-dashed lines are the ``thick'' atmosphere
and ``thin'' atmosphere with $\ycol=1.2\mbox{ g cm$^{-2}$}$,
respectively (see text for details).
The short-dashed horizontal line indicates the effective temperature
($\teff=5.3\times 10^5$~K) of the atmosphere model.
}
\label{fig:temp}
\end{figure}

\begin{figure}
\includegraphics[width=84mm]{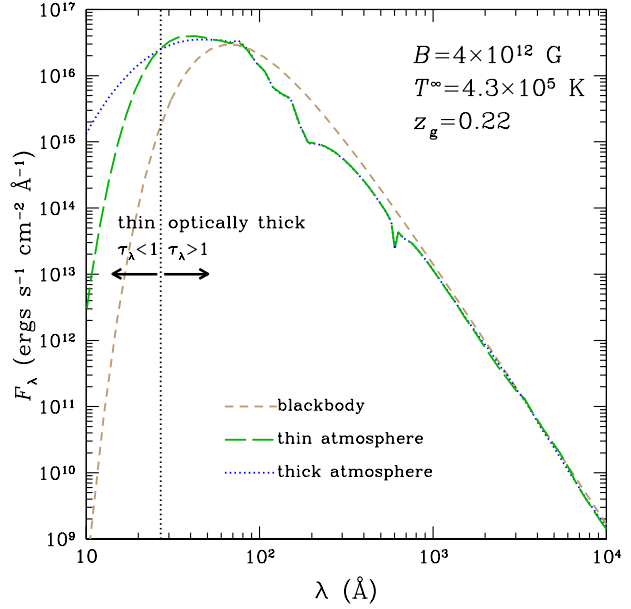}
\caption{
Spectra of hydrogen atmospheres with $B=4\times 10^{12}$~G and
$\tinfty=4.3\times 10^5$~K.
The dotted and long-dashed lines are the model spectra
using the ``thick'' atmosphere
and ``thin'' atmosphere with $\ycol=1.2\mbox{ g cm$^{-2}$}$,
respectively (see text for details).
The short-dashed line is for a blackbody with the same temperature.
All spectra are redshifted by $\zg=0.22$.
The vertical line separates the wavelength ranges where the atmosphere
is optically thin ($\taulam<1$) and optically thick ($\taulam>1$).
}
\label{fig:sp_thin}
\end{figure}

\begin{figure}
\includegraphics[width=84mm]{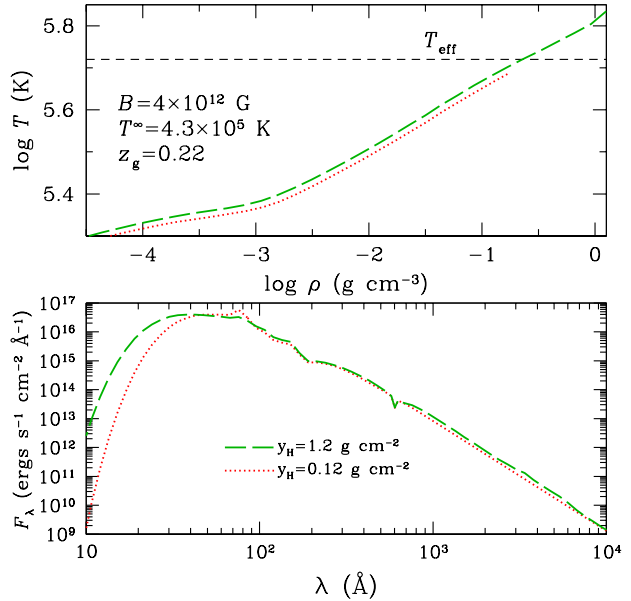}
\caption{
Model atmosphere temperature profiles (top) and spectra (bottom) with
$B=4\times 10^{12}$~G and $\tinfty=4.3\times 10^5$~K.
The long-dashed and dotted lines are the ``thin'' atmospheres with
$\ycol=1.2$ and $0.12\mbox{ g cm$^{-2}$}$, respectively.
The short-dashed horizontal line indicates the effective temperature
($\teff=5.3\times 10^5$~K) of the atmosphere model.
All spectra are redshifted by $\zg=0.22$.
}
\label{fig:thin2}
\end{figure}

The method for constructing a thin atmosphere does not differ significantly
from that used for a thick atmosphere (see Ho \& Lai~2001 for details on
thick atmosphere construction).  The difference is that, instead
of computing the model to large Thomson depths (where $\taulam \gg 1$
for all $\lambda$), we go to a relatively shallow depth at density
$\rho_{\rm thin}$.  At this shallow depth, we adjust the inner boundary
condition (depending on the model, either a blackbody or a condensed
surface spectrum; see Section~\ref{sec:cond}) to account for reflections.
The temperature profile is mainly determined by the radiative flux
(through radiative equilibrium) and mean opacities.
For instance, the profile is nearly the same in the thin and thick
atmosphere models shown in Figure~\ref{fig:temp} because the bulk
of the radiative energy comes from wavelengths for which the atmosphere is
optically thick; this is clearly seen in Figure~\ref{fig:sp_thin},
which plots the atmosphere spectra seen by a distant observer.
Thus, the difference in the spectra at short wavelengths basically does not
affect the mean (Rosseland-like) opacities and the resulting temperature
profile.  The only noticeable deviation occurs near the condensed surface
and is due to the boundary condition imposed at this depth.
For comparison, Figure~\ref{fig:thin2} shows a profile with still lower
$\rho_{\rm thin}$.  In this case, the boundary between the
optically thick and thin portions of the spectrum lies at a longer
wavelength.  Since the integrated flux is more sensitive to the spectral
difference at these longer wavelengths (being closer to the peak of the
spectrum), we see a more noticeable difference in the temperature profiles.
A qualitatively similar result is obtained for the same atmosphere
thickness but with a higher effective temperature, since the peak flux is
shifted to shorter wavelengths.
 
\subsection{Condensed Iron Versus Blackbody Emission}
\label{sec:cond}
 
In addition to atmosphere models in which the deepest layer
of the atmosphere is assumed to be a blackbody, we construct
(more realistic and self-consistent) models in
which this layer undergoes a transition from a gaseous
atmosphere to a condensed surface.
A surface composed of iron is a likely end-product of NS formation, and
Fe condenses at $\rho\approx 561\,AZ^{-3/5}B_{12}^{6/5}\mbox{ g cm$^{-3}$}
\approx 2.35\times 10^4\mbox{ g cm$^{-3}$}$ and
$T\la 10^{5.5}B_{12}^{2/5}\mbox{ K}\approx 5.5\times 10^5$~K for the
case considered here (Lai~2001); note that there is several tens
of percent uncertainty in the condensation temperature (Medin \& Lai,
private communication; see also Medin \& Lai~2006a,b).
The hydrogen condensation temperature is also uncertain (see, e.g.,
Lai \& Salpeter~1997; Potekhin, Chabrier, \& Shibanov~1999; Lai~2001);
the results of Lai~(2001) give $T\sim 2\times 10^4B_{12}^{0.65}$~K, which
suggests much lower temperatures than is relevant for our case.
The condensed matter surface possesses different emission properties
than a pure blackbody (Brinkmann~1980; Turolla, Zane, \& Drake~2004;
van Adelsberg \etal~2005;
P\'{e}rez-Azor\'{i}n, Miralles, \& Pons~2005); in particular, features can
appear at the plasma and proton cyclotron frequencies\footnote{
van Adelsberg \etal~(2005) consider an approximation in their treatment of
the ion contribution to the dielectric tensor which leads to a spectral
feature at the proton cyclotron frequency.  However, because of the
uncertainty in this approximation, the strength of the feature is not
well-determined.  Nevertheless, our results are not at all strongly
dependent on this feature (or the input spectrum at these low energies)
because the optical depth of the atmosphere $\taulam\gg 1$ at the
proton cyclotron (and plasma) frequency (see text for discussion).
\label{foot:condensed}}.

\begin{figure}
\includegraphics[width=84mm]{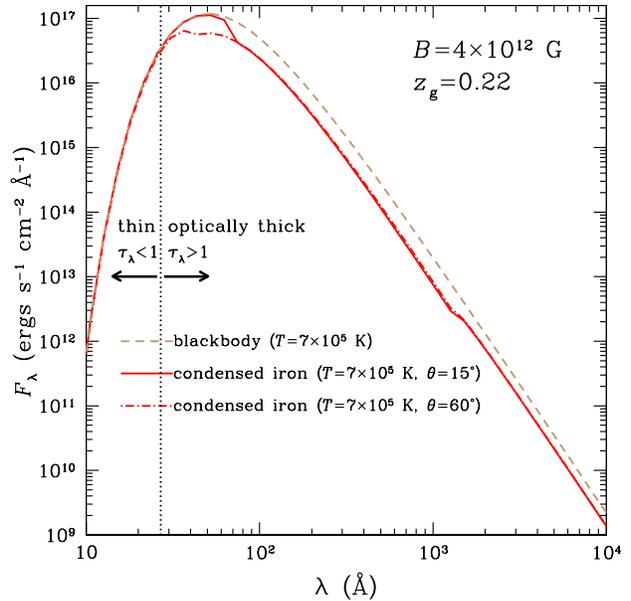}
\caption{
Condensed iron surface spectrum (solid line for photon propagation
direction $\theta=15^\circ$ and dot-dashed line for $\theta=60^\circ$),
with $B=4\times 10^{12}$~G, $T=7\times 10^5$~K, and
$\rho=2.35\times 10^4\mbox{ g cm$^{-3}$}$,
compared to a blackbody (dashed line) with the same temperature.
All spectra are redshifted by $\zg=0.22$.
The vertical line separates the wavelength ranges where the atmosphere
is optically thin ($\taulam<1$) and optically thick ($\taulam>1$).
}
\label{fig:sp_cond}
\end{figure}

We use the calculations of van Adelsberg \etal~(2005) to determine the
input spectrum
in our radiative transfer calculations of the atmosphere.  However,
at the temperature ($\tatm\approx 7\times 10^5$~K) of the condensed layer
relevant to our thin atmosphere models that fit the spectrum of \rxjw,
the input spectrum (where $\taulam\la 1$) is effectively unchanged from
a blackbody
(since the temperature profile is nearly identical,
with $\Delta\tatm\sim 3\%$; see Fig.~\ref{fig:temp}).
Thus there are only slight differences in the resulting surface spectrum.
This is illustrated in Figure~\ref{fig:sp_cond}, where we show the emission
spectrum from a $B=4\times 10^{12}$~G condensed iron surface at
$T=7\times 10^5$~K and $\rho=2.35\times 10^4\mbox{ g cm$^{-3}$}$
and compare this to a blackbody at the same temperature.
The deviation from a blackbody is smaller at low angles
(with respect to the magnetic field)
of photon propagation $\theta$ and increases
for increasing $\theta$, as illustrated by the two angles
$\theta=15^\circ$ and $60^\circ$ (see van Adelsberg \etal~2005).
Thus for most angles $\theta$, the condensed surface
spectra at short wavelengths (where $\taulam\ll 1$, so that this surface
is visible
to an observer above the atmosphere) are virtually identical to a blackbody.
On the other hand, the atmosphere is optically thick at longer wavelengths,
where the condensed surface spectra deviate from a blackbody; thus
the condensed surface and the spectral features are not visible.
We see from Figure~\ref{fig:sp_thin} that the
harder spectrum at high energies in the ``thick'' atmosphere becomes much
softer in the ``thin'' atmosphere and takes on a blackbody shape.
In contrast, there is a negligible difference where the atmosphere is
optically thick.

\section{Observations \& Analysis}
\label{sec:obs}

We collect publically available optical, UV, and X-ray data on
\rxjw.  These data have been discussed elsewhere so our treatment will
be brief.  First, we assemble the optical ($B$- and $R$-band)
photometry from the Very Large Telescope (VLT) from van Kerkwijk \&
Kulkarni~(2001a) and the \textit{Hubble Space Telescope} (\hst) WFPC2
F170W, F300W, F450W, and F606W photometry (Walter~2001; Pons
\etal~2002) as analyzed by van Kerkwijk \& Kulkarni~(2001a).  We also
take the optical VLT spectrum from van Kerkwijk \& Kulkarni~(2001a) and
a STIS far-UV spectrum.\footnote{datasets: O5G701010-O5G701050,
O5G702010-O5G702050, O5G703010-O5G703050, O5G704010-O5G704050,
O5G705010-O5G705050, O5G751010-O5G751050, and O5G752010-O5G752050.}
The spectra are entirely consistent with the photometry as calibrated
by van Kerkwijk \& Kulkarni~(2001a), although given the limited
signal-to-noise ratio of the spectra, we rely primarily on the
photometry in what follows.  We then use the \textit{Extreme
Ultraviolet Explorer} (\euve; Haisch, Bowyer, \& Malina~1993) data as
discussed and reduced by Pons \etal~(2002).  Finally, we take the RGS
spectrum from the 57-ks \xmm\ observation and the 505-ks \chandra\
LETG spectrum that are discussed by Burwitz \etal~(2003).

A source of uncertainty is that, as mentioned by Burwitz
\etal~(2003), the RGS and LETG data are not entirely consistent in
terms of flux calibration: while they have very similar shapes (and
hence implied temperatures and absorptions) the radii inferred from
blackbody fits differ by as much as 10\% and the overall flux by as
much as 20\%.
Since the LETG fits in Burwitz \etal~(2003) are more
consistent with those of the CCD instruments on \xmm\ (EPIC-pn and
EPIC-MOS2) and in our opinion the current low-energy calibration of
LETG is more reliable, we adjust the flux of the RGS data upward by
17\% to force agreement with the \chandra\ data.  We do not know for
certain which calibration (if either) is entirely accurate, so some
care must be taken when interpreting the results at the 10\%--20\%
level.  Fully reliable calibration or even cross-calibration at the
low-energy ends of the \chandra\ and \xmm\ responses
($<0.2$~keV) is not currently available (see, e.g., Kargaltsev
\etal~2005), and the detailed response of \euve\ compared to those
of \chandra\ and \xmm\ is also unknown.
Therefore, for accuracy in doing the EUV/X-ray fits, we concentrate
on the LETG data, which are consistent and have high-quality
calibration.

We follow the HRC-S/LETG analysis threads ``Obtain
Grating Spectra from LETG/HRC-S Data''\footnote{See
http://asc.harvard.edu/ciao/threads/spectra\_letghrcs/.},
``Creating Higher-order Responses for HRC-S/LETG
Spectra''\footnote{See
http://asc.harvard.edu/ciao/threads/hrcsletg\_orders/.},
``Create Grating RMFs for HRC Observations''\footnote{See
http://asc.harvard.edu/ciao/threads/mkgrmf\_hrcs/.}, and
``Compute LETG/HRC-S Grating ARFs''\footnote{See
http://asc.harvard.edu/ciao/threads/mkgarf\_letghrcs/.} and
use \texttt{CIAO}\footnote{
http://cxc.harvard.edu/ciao/}
version 3.2.2 and \texttt{CALDB} version 3.2.2.
We extracted the dispersed events and generated response files for
orders $\pm 1$, $\pm 2$, and $\pm 3$.
After fitting the LETG data,
we compare the fit results qualitatively with the RGS and
\textit{EUVE} data; the general agreement is good, but we do not use
them quantitatively.

\section{Atmosphere Model Fitting}
\label{sec:atmfit}

Because of data reduction and cross-calibration issues (see
Section~\ref{sec:obs}) and possible variations in the interstellar
absorption abundances (standard abundances are assumed), we do not feel that
a full fit of the data is justified at this time.  Therefore we do not
fit for all of the parameters in a proper sense nor do we perform a
rigorous search of parameter space.  Instead, we fit for a limited
subset of parameters while varying others manually.  This allows us to
control the fits in detail and reduce the computational burden of
preparing a continuous distribution (in $B$, $\teff$, and $\ycol$)
of models, yet still determine whether our model qualitatively
fits the data.
 
\begin{figure}
\includegraphics[width=84mm]{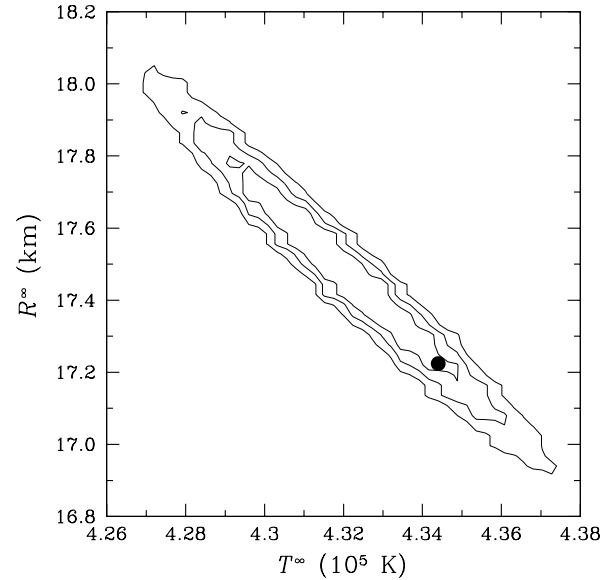}
\caption{
1, 2, 3$\sigma$ confidence regions for the fit parameters
$\tinfty$ and $\rinfty$ of the $B=4\times 10^{12}$~G atmosphere model.
The $3\times 10^{12}$~G model gives very similar results.
The dot indicates the best-fit point given in Table~\ref{tab:fitparam}.
}
\label{fig:b4fit}
\end{figure}

For a given magnetic field and atmosphere thickness, we generate partially
ionized atmosphere models for a range of effective temperatures.
(Note that, since the continuum opacity of the dominant photon polarization
mode decreases for increasing magnetic fields, the required thickness
$\ycol$ of the atmosphere increases for increasing $B$.)
We then perform a $\chi^2$ fit in \texttt{CIAO} to the LETG data over
the 10--100~\AA\ range (0.12--1.2~keV) for the absorption column
density \nh\ [using the \texttt{TBabs} absorption model from
\texttt{Xspec} (Wilms, Allen, \& McCray 2000), although other models
such as \texttt{phabs} give similar results], the temperature \tinfty,
the normalization (parameterized by \rinfty), and the redshift $\zg$,
where we interpolate over \tinfty.
We obtain a good fit, and Table~\ref{tab:fitparam} lists the best-fit
parameters and models;
the radius is given assuming a distance $d_{140}=d/(140\mbox{ pc})=1$
(see footnote~\ref{foot:distance}).
Figure~\ref{fig:b4fit} shows the confidence regions for temperature and
radius; the covariance between the other fit parameters are not as
significant.
While we find a range of magnetic fields that give acceptable fits,
changes in the magnetic field outside this range (but still within
$B=10^{12}-10^{13}$~G) and atmosphere thickness lead to worse fits.
At $B > 10^{13}$~G, spectral features due to proton cyclotron and
bound species appear in the observable soft X-ray range
(though they are likely to be broadened; see Section~\ref{sec:pulse}),
which are not seen.

\begin{table}
\begin{minipage}{8cm}
\caption{Fits to the X-ray Data \label{tab:fitparam}}
\begin{tabular}{l c c c}
\hline
 & \mc{2}{c}{Atmosphere} & Blackbody \\
\hline
 & \mc{2}{c}{Model Parameters} & \\ \\
$B$ ($10^{12}$) & 3 & 4 & \\
$\ycol$ (g cm$^{-2}$) & 1.2 & 1.2 & \\
\\ \mc{4}{c}{Fit Results
 \footnote{Numbers in parentheses are 68\% confidence limits in the last
digit(s).}
 } \\ \\
\nh ($10^{20}\mbox{ cm}^{-2}$) & 1.18(2) & 1.30(2) & 0.91(1) \\
\tinfty ($10^5$~K) & 4.34(3) & 4.34(2) & 7.36(1) \\
\rinfty ($d_{140}$~km)\footnote{The formal fit uncertainty is $< 10\%$.
However, since the radius determination depends on the distance, we
conservatively adopt the current $\sim 30\%$ distance uncertainty
as our radius uncertainty.}
 & 17 & 17 & 5.0 \\
$\zg$ & 0.27(2) & 0.22(2) & \\
$\chi_{\rm r}^2$/dof & 0.85/4268 & 0.86/4268 & 0.86/4269 \\
\hline
\end{tabular}
\end{minipage}
\end{table}

\begin{figure}
\includegraphics[width=84mm]{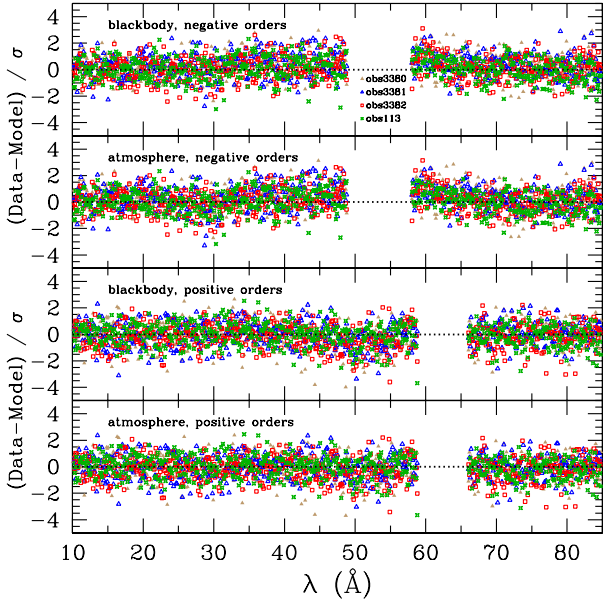}
\caption{
Fit residuals for the blackbody and $B=4\times 10^{12}$~G atmosphere
model given in Table~\ref{tab:fitparam}.
}
\label{fig:resid}
\end{figure}

To further evaluate the quality of this fit, we fit the same LETG data
with a blackbody.  The blackbody fit yields parameters
(see Table~\ref{tab:fitparam}) that are very
similar to those derived by Burwitz \etal~(2003; see
Section~\ref{sec:intro}).  A comparison of the residuals for the
blackbody fit and the $B=4\times 10^{12}$~G atmosphere model fit is
plotted in Figure~\ref{fig:resid}.  Given the comparable
$\chi_{\rm r}^2$ ($\approx 1$) we achieve from our blackbody and
atmosphere model fits (along with the low-energy calibration
uncertainties), we are confident that the atmosphere model describes
the observations just as well as a blackbody.

Next, we examine the quality of the fit to the optical/UV data.  van
Kerkwijk \& Kulkarni~(2001a) showed that these data are well fit by a
Rayleigh-Jeans power law ($F_{\lambda} \propto \lambda^{-4}$) with an
extinction $A_{V}=0.12 \pm 0.05$~mag.  In our fitting, we try using
both the optical extinction implied by the X-ray absorption
[$A_{V}=\nh/(\expnt{1.79}{21}\mbox{ cm}^{-2})$~mag; Predehl \&
Schmitt~1995] and fitting freely for $A_V$, but we find that these
fits are too unconstrained and that the value of $A_V$ is not
sensitively determined by the data (indeed, this is reflected in the
large uncertainties found by van Kerkwijk \& Kulkarni~2001a).  As a
result, we fix $A_V$ to 0.12~mag.
We also assume a single value for the reddening ($R_{V}=3.1$) and use the
reddening model of Cardelli, Clayton, \& Mathis~(1989) with
corrections from O'Donnell~(1994).  We find that our best-fit model to
the X-ray data also produces a $\lambda^{-4}$ power law but
underpredicts the optical/UV data by a factor of 15$-$20\%\footnote{
Comparing monochromatic fluxes derived from
photometry with the models is not sufficiently accurate for a
detailed, quantitative analysis.  A more accurate method would involve
convolving the models with the filter bandpasses and predicting
monochromatic count-rates to compare with the data (e.g., van Kerkwijk
\& Kulkarni~2001a; Kaplan \etal~2003).  However, given the assumptions
about extinction and reddening and the level of accuracy of this
analysis, the first approach will suffice here.}.  Looking in detail
at the highest quality optical data point (the \hst\ F606W measurement),
we find that it is 15\% above our model spectra.  However, the error budget
is 3\% (photometric uncertainty), 5\% ($A_V$ uncertainty),
10\% (uncertainty in the optical model flux), and 5\% (uncertainty
in the fitted optical flux due to the X-ray normalization), and therefore
the 15\% disagreement can easily be explained by known sources of
uncertainty.

\begin{figure}
\includegraphics[width=84mm]{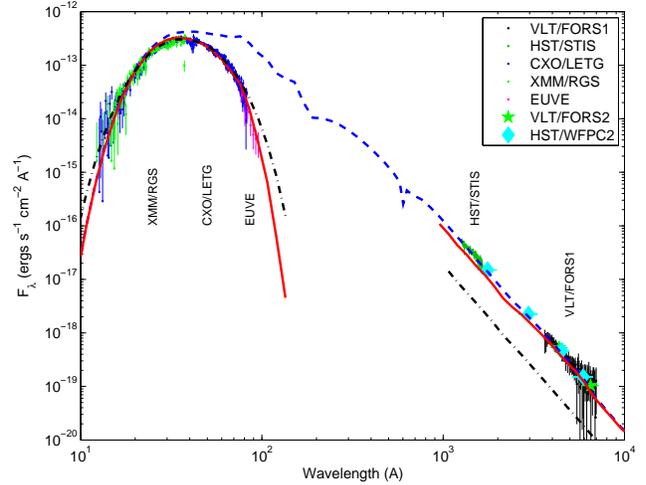}
\caption{
Spectrum of \rxjw\ from optical to X-ray wavelengths.
The data points are observations taken from various sources.
Error bars are one-sigma uncertainties.
Optical spectra are binned for clarity: STIS data into 30 bins at a
resolution of 12~\AA\ and VLT data into 60 bins at 55~\AA\ resolution.
The solid line is the absorbed (and redshifted by $\zg=0.22$) atmosphere
model spectrum with $B=4\times 10^{12}$~G, $\ycol=1.2\mbox{ g cm$^{-2}$}$,
$\tinfty = 4.3\times 10^5$~K, and $\rinfty=17$~km.
The dashed line is the unabsorbed atmosphere model spectrum.
The dash-dotted line is the (absorbed) blackbody fit to the X-ray spectrum
with $\rinfty=5$~km.
}
\label{fig:sp}
\end{figure}

Figure~\ref{fig:sp} shows the observations of \rxjw.
We also overlay the blackbody and our $B=4\times 10^{12}$~G atmosphere
model spectra with the parameters given in Table~\ref{tab:fitparam}.
As discussed in Section~\ref{sec:intro}, the data are generally featureless,
while the models show spectral features;
at $B\approx (3-4)\times 10^{12}$~G, the features due to bound species
lie in the extreme UV to very soft X-ray range and are thus hidden by
interstellar absorption.
Overall, we see that our atmosphere model spectra are
generally consistent with the X-ray and optical/UV data, while a blackbody
underpredicts the optical/UV by a factor of 6$-$7.

\section{Discussion}
\label{sec:discussion}
Two important uncertainties in our model for \rxjw\ are discussed here.
The first is flux variability as a result
of surface magnetic field and temperature distributions.
The second is the creation of thin hydrogen atmospheres.
 
\subsection{Surface Magnetic Field and Temperature Variations and
Constraints on Pulsations}
\label{sec:pulse}

A major unexplained aspect of \rxjw\ is the strong observational
limit on the lack of pulsations ($\la 1.3\%$).
As discussed by Braje \& Romani~(2002), the low pulsations could be
due to an unfavorable viewing geometry and/or a very close alignment
between the rotation and magnetic axes.
Assuming blackbody emission, they find this is possible in
$\sim 5\%$ of viewing geometries (with $R=14$~km).
In fact, they note that if \rxjw\ is a normal radio pulsar, then the lack
of a radio detection may imply low X-ray pulsations, since the pulsar beam
does not cross our line of sight.
On the other hand, atmosphere emission can be highly beamed in the
presence of a magnetic field (Shibanov \etal~1992; Pavlov \etal~1994)
and magnetic field variations over the NS surface will induce surface
temperature variations (Greenstein \& Hartke~1983).
There arises then the questions of whether our single temperature and
magnetic field (strength and orientation) atmosphere model fit is
physically correct and whether it produces stronger pulsations than
is observed.  In response to the former,
clearly our model spectrum represents an average over the surface and
the single temperature and magnetic field indicate ``mean'' values.

In order to determine the strength of pulsations, synthetic spectra
from the whole NS surface must be calculated.  Such synthetic spectra are
necessarily model-dependent (see, e.g., Zane \etal~2001; Ho \& Lai~2004;
Zane \& Turolla~2005, 2006), as the magnetic field and temperature
distributions over the NS surface are unknown.
Therefore, we adopt a relatively simple model:
we assume the surface is symmetric (in $B$ and $T$) about the magnetic
equator and divide the hemisphere into four magnetic latitudinal regions.
We generate atmosphere models for each region with the parameters given
in Table~\ref{tab:nssurface}
(note that the magnetic field distribution is roughly dipolar and
$\ThetaB$ is the angle between the magnetic field and the surface normal).
Using an analogous formalism to that described in Pechenick \etal~(1983)
and Pavlov \& Zavlin~(2000), we calculate phase-resolved spectra and
light curves from the whole NS surface
(we assume $M=1.4 M_\odot$ and $R=14$~km).
The resulting phase-resolved spectra have bound and cyclotron features
that are broadened due to the surface magnetic field and temperature
variations but otherwise are not significantly different from the
``local'' spectrum we use in Section~\ref{sec:atmfit};
thus they would produce a qualitatively similar fit.
From the light curves, we obtain energy-integrated pulse fractions
for the wavelength range 10$-$85~\AA\ (the approximate range of the
X-ray observations).

\begin{table}
\caption{Parameters for Neutron Star Surface \label{tab:nssurface}}
\begin{tabular}{c c c c}
\hline
magnetic latitude & $B$ & $\teff$ & $\ThetaB$ \\
 (deg) & ($10^{12}$~G) & ($10^5$~K) & (deg) \\
\hline
0---10 & 6 & 7 & 0 \\
10---40 & 5 & 6 & 30 \\
40---70 & 4 & 5 & 60 \\
70---90 & 3 & 4 & 90 \\
\mc{4}{c}{or} \\
70---90 & - & 0 & - \\
\hline
\end{tabular}
\end{table}

Assuming the magnetic axis is orthogonal to the rotation axis,
$f_{\rm P}< 0.01, 0.04, 0.07$ for $\zeta=10^\circ, 20^\circ, 30^\circ$,
respectively, where
$f_{\rm P}=(F_{\rm max}-F_{\rm min})/(F_{\rm max}+F_{\rm min})$
is the pulse fraction and
$\zeta$ is the angle between the rotation axis and the direction to
the distant observer.
Extending the arguments of Braje \& Romani~(2002) to realistic atmosphere
emission, we thus find there exist viewing geometries for which the
pulse fraction is at the observed limits, despite the relatively large
magnetic field (and hence strong beaming) and temperature variations.
We also note the following:
(1) An H$\alpha$ nebula is found around \rxjw; if interpreted as a bow shock
that is powered by the rotational energy loss of a NS magnetic dipole,
then the spin period $P\approx 0.5(B/10^{12}\mbox{ G})^{1/2}$~s
(van Kerkwijk \& Kulkarni~2001b), so that $P\sim 1$~s.
(2) The emission can be strongly beamed at higher energies, so that
the pulse fraction is energy-dependent; pulsations may be more
evident at the higher X-ray energies, although the soft spectrum of
\rxjw\ means that broad energy ranges are likely still the most sensitive
to pulsations.

\subsection{Creation of Thin Hydrogen Atmospheres}
\label{sec:dnb}
 
From the atmosphere model fits in Section~\ref{sec:atmfit}
(see also Fig.~\ref{fig:sp_thin})
the critical wavelength $\lambatm\approx 30$~\AA\
gives an atmosphere column $\ycol\approx 1-2\mbox{ g cm$^{-2}$}$.
A hydrogen atmosphere with this column density has a
total mass of $M_{\rm H}\approx 1\times 10^{13}(R/10\mbox{ km})^2
(\ycol/1\mbox{ g cm$^{-2}$})$~g or about $10^{-21}$ of
the mass of a $1.4\,M_{\odot}$ NS.  Since the diffusion timescale
is extremely short in a high gravity environment (Alcock \& Illarionov~1980),
the atmosphere contains the bulk of the total hydrogen budget of the NS.
The origin of such thin H layers is a problem which we now address by
briefly discussing possible mechanisms for generating thin H atmospheres.
We leave a more detailed study to future work.

Accretion at low rates may create a thin hydrogen layer on the
NS surface.  A thin hydrogen layer of $\ycol = 1\,{\rm g\ cm}^{-2}$
has a total mass of $\approx 6\times 10^{-21}\ M_{\odot}$.  Hence the
time-averaged accretion rate over the age of \rxjw\ ($\sim 5 \times 10^5$~yr;
see Section~\ref{sec:conclusion}) is $\approx 0.8\,{\rm g\ s}^{-1}$.
For the case of RX~J0720.4$-$3125 with $\ycol=0.16\mbox{ g cm$^{-2}$}$,
such a mass layer may result if the time-averaged accretion rate onto
the NS is about $0.06\mbox{ g s$^{-1}$}$ over $10^6$~yr.
Such small amounts of accretion indicate an enormous fine-tuning
problem.  Thus this process is unlikely to be the origin of the thin
hydrogen envelope.
 
We next consider the possibility that the thin hydrogen layer is a
remnant from the formation of the NS.  In this case,
Chang \& Bildsten~(2003, 2004) point out that diffusive nuclear burning (DNB)
is the dominant process for determining what happens to hydrogen on the
surface of a NS.
Due to the power law dependence of the column lifetime $\tau_{\rm col}$
on $\ycol$, DNB leaves a thin layer of hydrogen, whose thickness depends
on the thermal history of the NS and the underlying composition.
We find the column lifetime from a full numerical solution to be
$\tau_{\rm col} \approx 10^{10}$~yr for a magnetized envelope with
$B=10^{12}$ G and $\tau_{\rm col}\approx 10^8$~yr for $B=10^{13}$ G
(assuming $M=1.4\,M_\odot$, $R=14$~km, $\teff=5\times 10^5$~K,
and $\ycol=1\mbox{ g cm$^{-2}$}$).
The shorter column lifetime for higher magnetic fields is due to
the effect of the magnetic field on the electron equation of state, 
which increases the hydrogen number density in the burning layer
(Chang, Arras, \& Bildsten~2004).
Thus there is insufficient time at the current effective temperature to
reduce the surface hydrogen to such a thin column, assuming that the age
of \rxjw\ is $\sim 5 \times 10^5$~yr.
However, DNB was much more effective in the past when the NS was hotter.  
In fact, during the early cooling history of the NS, DNB would have
rapidly consumed all the hydrogen on the surface
(Chang \& Bildsten 2004).
This leads to another fine-tuning problem: if we fix
the lifetime of an envelope to some value representing the time the NS
spends at a particular temperature, we find that the resulting
column scales like $\ycol\propto \teff^{-40}$ (Chang \& Bildsten~2003);
therefore, the size of the atmosphere is extremely sensitive to the early
cooling history of the NS.  As a result, DNB appears to be a less likely
mechanism for producing such thin atmospheres.

Finally, we briefly examine a self-regulating mechanism that is driven by
magnetospheric currents and may produce thin hydrogen layers on the
surface.  Since the accelerating potential could be as high as 10~TeV
(Arons~1984), high energy particles would create electromagnetic
cascades on impact with the surface (Cheng \& Ruderman~1977).  These
cascades result in $e/\gamma$ dissociation of surface material,
analogous to proton spallation of CNO elements in accreting systems (Bildsten,
Salpeter, \& Wasserman~1992).  Protons would be one of the products of
this dissociation and would rise to the surface due to the rapid
diffusion timescale.  Since protons cannot further dissociate into
stable nuclei, a hydrogen layer is built up to a column roughly given
by the radiation length of the ultrarelativistic electrons, beyond
which hydrogen can no longer be produced and the above mechanism
terminates.  The radiation length of ultrarelativistic electrons
depends on the stopping physics. In the zero magnetic field limit, the
stopping physics is dominated by relativistic bremsstrahlung, and the
cross-section is $\sigma\sim\alpha_{\rm F} Z^2 r_0^2$, where
$\alpha_{\rm F}\equiv e^2/\hbar c$ is the fine structure constant, $Z$
is the charge number of the nuclei, and $r_0=e^2/m_{\rm e}c^2$ is the
classical electron radius (Bethe \& Heitler~1934; Heitler~1954).  This
gives a stopping column $y_{\rm stop}\sim 3000$~g~cm$^{-2}$ for
hydrogen.  Bethe \& Ashkin~(1953) and Tsai~(1974) give a more accurate
value, $y_{\rm stop}\sim 60$~g~cm$^{-2}$, for the stopping column of
atomic hydrogen; thus it appears that the resulting columns are too
thick.  However, for sufficiently strong magnetic fields, the stopping
physics of ultrarelativistic electrons is significantly modified, and
the dominant stopping mechanism is via magneto-Coulomb interactions
(Kotov \& Kel'ner~1985; see also Kotov, Kel'ner, \& Bogovalov~1986).
In magneto-Coulomb stopping, relativistic electrons traveling along
field lines are kicked up to excited Landau levels via collisions and
de-excite, radiating photons with energy $\gamma\hbar\omega_B$, where
$\hbar\omega_B$ is the Landau cyclotron energy and $\gamma$ is the
Lorentz factor of the incoming electron.  Compared to zero-field
relativistic bremsstrahlung, the magneto-Coulomb radiation length is
smaller by a factor of $\sim\pi\alpha_{\rm F}$ (Kotov \& Kel'ner
1985).  Applying the correction to the Bethe \& Ashkin~(1953) result
($y_{\rm stop}\sim 60$~g~cm$^{-2}$), we find this gives roughly the
required thin atmosphere column, $y_{\rm stop}\sim 1$~g~cm$^{-2}$.
Though extremely suggestive, this mechanism requires a more complete
study and is the subject of future work
(see, e.g., Thompson \& Beloborodov~2005; Beloborodov \& Thompson~2006).

\section{Summary and Conclusions} \label{sec:conclusion}

We have gathered together observations of the isolated neutron
star \rxjw\ and compared them to our latest magnetic, partially
ionized hydrogen atmosphere models.
Prior works showed that the observations were well-fit by blackbody spectra.
Here we obtain good fits to the overall multiwavelength spectrum
of \rxjw\ using the more realistic atmosphere model.
In particular, we do not overpredict the optical flux obtained by
previous works and require only a single temperature atmosphere.
[Note that this single temperature (and magnetic field) serves as an
average value for the entire surface.]
In addition to the neutron star orientation and viewing geometry, the
single temperature helps to explain the non-detection of pulsations thus far.
At high X-ray energies, where the atmosphere is optically thin,
the model spectrum has a ``blackbody-like'' shape due to the emission
spectrum of a magnetized, condensed surface beneath the atmosphere.
The atmosphere is optically thick at lower energies;
thus features in the emission spectrum of the condensed surface are not
visible when viewed from above the atmosphere.
The ``thinness'' of the atmosphere helps to produce the featureless,
blackbody-like spectrum seen in the observations.
Using a simple prescription for the temperature and magnetic field
distributions over the neutron star surface,
we obtain pulsations at the currently observed limits.

Based on a possible origin within the Upper Scorpius OB association,
the age of \rxjw\ is estimated to be about $5\times 10^5$~yr
(Walter~2001; Walter \& Lattimer~2002; Kaplan \etal~2002).
Our surface temperature determination ($k\tinfty=37$~eV) is only
a factor of 1.7
below the blackbody temperature ($k\tinfty=63$~eV) obtained by previous
works (see Sections~\ref{sec:intro} and \ref{sec:atmfit}) and therefore
does not place much stronger constraints on theories of neutron star
cooling (see, e.g., Page \etal~2004; Yakovlev \& Pethick~2004).
It may also be noteworthy that \rxjw\ is one of the cooler isolated
neutron stars and possibly possesses the lowest magnetic field
[$B\approx (3-4)\times 10^{12}$~G]; the lower magnetic field implies a more
uniform surface temperature distribution and weaker radiation beaming.

We show that the production of thin hydrogen layers
is difficult via accretion or diffusive nuclear burning.
Accretion requires a time-averaged rate that is extremely small and
fine-tuned.  Diffusive nuclear burning has no effect at the current
effective temperature but is very efficient when the star was hotter,
such that the column thickness depends very sensitively on the early cooling
history of the neutron star.
Hence these scenarios cannot easily explain the thin atmosphere columns.
On the other hand, it may be possible for magnetospheric currents to produce
the hydrogen atmospheres if the stopping physics of relativistic
electrons is dominated by magneto-Coulomb interactions.
We note that younger neutron stars may possess
atmospheres composed of helium rather than hydrogen;
however, the opacities of bound states of helium in strong magnetic
fields is still largely unknown (see, e.g., Al-Hujaj \& Schmelcher~2003a,b;
Bezchastnov, Pavlov, \& Ventura~1998; Pavlov \& Bezchastnov~2005).

Finally, the emission radius we derive from our atmosphere model fits is
$\rinfty\approx 17\,(d/140\mbox{ pc})$~km
(although recall the distance and flux uncertainties discussed in
footnote~\ref{foot:distance} and Section~\ref{sec:obs}, respectively).
Accounting for gravitational redshift ($\zg\sim 0.22$),
this yields $\rem\approx 14$~km.
This is much larger than the inferred radius obtained by just fitting the
X-ray data with a blackbody
($\rinfty_{\rm X}\approx 5$~km; see Sections~\ref{sec:intro} and \ref{sec:atmfit}).
As a result, there does not appear to be a need to resort to more exotic
explanations such as quark or strange stars (e.g., Drake \etal~2002; Xu~2003;
see Walter~2004; Weber~2005 for a review), at least for the case of \rxjw.
On the other hand, our radius is small compared to the radius derived from
fitting the optical/UV data ($\rinfty_{\rm opt}\approx 21$~km).
For a $1.4\,M_\odot$ neutron star, the latter implies a low redshift
($\zg\approx 0.12$) and very large intrinsic radius
($\rem_{\rm opt}\approx 19$~km);
this is ruled out by neutron star equations of state,
while our radius $\rem\approx 14$~km only requires a standard, stiff
equation of state (see, e.g., Lattimer \& Prakash~2001).

\section*{Acknowledgments}

We greatly appreciate the useful discussion and observational data provided
to us by Marten van Kerkwijk.
We are very grateful to Gilles Chabrier for help on the opacity tables.
We thank Chris Thompson for discussion concerning the stopping physics
of relativistic electrons in strongly magnetized plasmas and for
pointing out the possible implications of currents along closed field
lines.
We thank Dong Lai for pointing out that magnetospheric currents can cause
nuclear reactions on the surface of active pulsars and for very helpful
discussions of their microphysics during the course of this work.  We
thank Lars Bildsten for enlightening conversations on the nature of
optical emission from INSs.
We thank the organizers of the NATO-ASI June 2004 meeting ``Electromagnetic
Spectrum of Neutron Stars,'' where some of the initial ideas for this work
were developed.
W.H. is supported by NASA through Hubble Fellowship grant HF-01161.01-A
awarded by STScI, which is operated by AURA, Inc., for NASA, under
contract NAS~5-26555.
D.K. was partially supported by a fellowship from the Fannie and John Hertz
Foundation and by HST grant GO-09364.01-A.
Support for the HST grant GO-09364.01-A was provided by NASA through a grant
from STScI, which is operated by AURA, Inc., under NASA contract NAS~5-26555.
P.C. is supported by the NSF under PHY~99-07949, by JINA through NSF grant
PHY~02-16783, and by NASA through Chandra Award Number GO4-5045C issued by
CXC, which is operated by the SAO for and on behalf of NASA under
contract NAS~8-03060.
P.C. acknowledges support from the Miller Institute for Basic Research
in Science, University of California Berkeley.
A.P. is supported by FASI through grant NSh-9879.2006.2 and by RFBR
through grants 05-02-16245 and 05-02-22003.


\label{lastpage}

\end{document}